# Application of Machine Learning in Melanoma Detection and the Identification of 'Ugly Duckling' and Suspicious Naevi: A Review

*Fatima Al Zegair[1], Nathasha Naranpanawa[1], Brigid Betz-Stablein[2], Monika Janda[3], H. Peter Soyer[2] and Shekhar S. Chandra[1]*

[1] School of Electrical Engineering and Computer Science, University of Queensland, Brisbane, QLD, Australia

[2] Frazer Institute, The University of Queensland, Dermatology Research Centre, Brisbane, QLD, Australia

[3] Centre for Health Services Research, Faculty of Medicine, The University of Queensland, Brisbane, QLD, Australia

## Abstract

Skin lesions known as naevi exhibit diverse characteristics such as size, shape, and colouration. The concept of an "Ugly Duckling Naevus" comes into play when monitoring for melanoma, referring to a lesion with distinctive features that sets it apart from other lesions in the vicinity. As lesions within the same individual typically share similarities and follow a predictable pattern, an ugly duckling naevus stands out as unusual and may indicate the presence of a cancerous melanoma. Artificial intelligence (AI) involves the use of computer systems to imitate intelligent behaviour, aiming to minimize human intervention. Computer-aided diagnosis (CAD) has become a significant player in the research and development field, as it combines machine learning techniques with a variety of patient analysis methods. Its aim is to increase accuracy and simplify decision-making, all while responding to the shortage of specialized professionals. These automated systems are especially important in skin cancer diagnosis where specialist availability is limited. As a result, their use could lead to life-saving benefits and cost reductions within healthcare. Given the drastic change in survival when comparing early stage to late-stage melanoma, early detection is vital for effective treatment and patient outcomes. Machine learning (ML) and deep learning (DL) techniques have gained popularity in skin cancer classification, effectively addressing challenges, and providing results equivalent to that of specialists. Despite these advancements, reviews on ML and DL approaches for identifying suspicious naevi and ugly duckling (UD) naevus are limited. This article provides an extensive overview of cutting-edge ML and DL-based algorithms for melanoma detection and the identification of suspicious naevi and UD naevus. The article commences with general information on skin cancer, melanoma, and various naevus types. It then presents an overview of AI, ML, DL, and CAD, followed by successful applications of different ML techniques including convolutional neural networks (CNN) for melanoma detection, comparing them with dermatologists' performance. Lastly, the article discusses ML methods for UD naevus detection and the identification of suspicious naevi.

## 1. Introduction

Skin cancer is recognized as a prevailing and significant form of human malignancy, exhibiting a considerable global burden in terms of incidence rates [1]. Skin cancer cells are categorized into different types, including Basel cell carcinoma (BCC), Squamous cell carcinoma (SCC), and melanoma [2].Visual examination stands as the primary modality employed in the diagnostic process, involving an initial clinical screening to detect potential skin cancer lesions. Subsequently, a dermoscopic analysis may be employed to further evaluate these lesions, while biopsy and subsequent histopathological examination serve as confirmatory measures if required for definitive diagnosis.

Melanoma, as an individual category within the spectrum of skin cancers, assumes exceptional significance, making a substantial contribution to the mortality rates associated with this disease. Its status as the most deadly subtype of skin cancer stems from its aggressive characteristics, characterized by rapid dissemination to distant anatomical sites when timely detection and treatment are lacking [3]. Exposure to both natural and artificial ultraviolet (UV) radiation is widely recognized as the primary causative factor for all types of skin cancer, including melanoma [4].

Melanoma commonly presents as a newly developed or changing melanocytic lesion on the skin, exhibiting notable characteristics such as an irregular shape, atypical coloration, and variable size. These distinctive attributes serve as differentiating factors from other types of skin lesions. Nonetheless, the distinction between melanoma and benign skin lesions can occasionally pose challenges, necessitating the utilization of specialized diagnostic tools and techniques to enhance diagnostic accuracy. [3]. Although constituting only 4% of skin cancer incidents, malignant melanoma is responsible for roughly 75% of mortality cases associated with skin cancer [5]. Lastly, The timely detection of melanoma is of critical significance due to its potential for cure in up to 95% of cases when promptly identified and appropriately treated. [6].

A naevus, the clinical term for a mole, is a benign tumor consisting of melanocytes that manifests as small brown, tan, or pink spots on the skin surface [7] . Naevi exhibit significant variation in their morphological characteristics, including diverse shapes, sizes, and colours [7]. Having many melanocytic naevi has been identified as a risk factor for melanoma, with an increased number of naevi being associated with a higher risk of developing melanoma [8, 9]. Studies indicate that the quantity, distribution, and dimensions of melanocytic naevi are influenced by environmental and behavioural factors associated with ultraviolet (UV) radiation exposure, in addition to genetic elements [10]. Figure 1 illustrates examples of suspicious and non-suspicious melanocytic naevi as well as other types of skin lesion cancers.

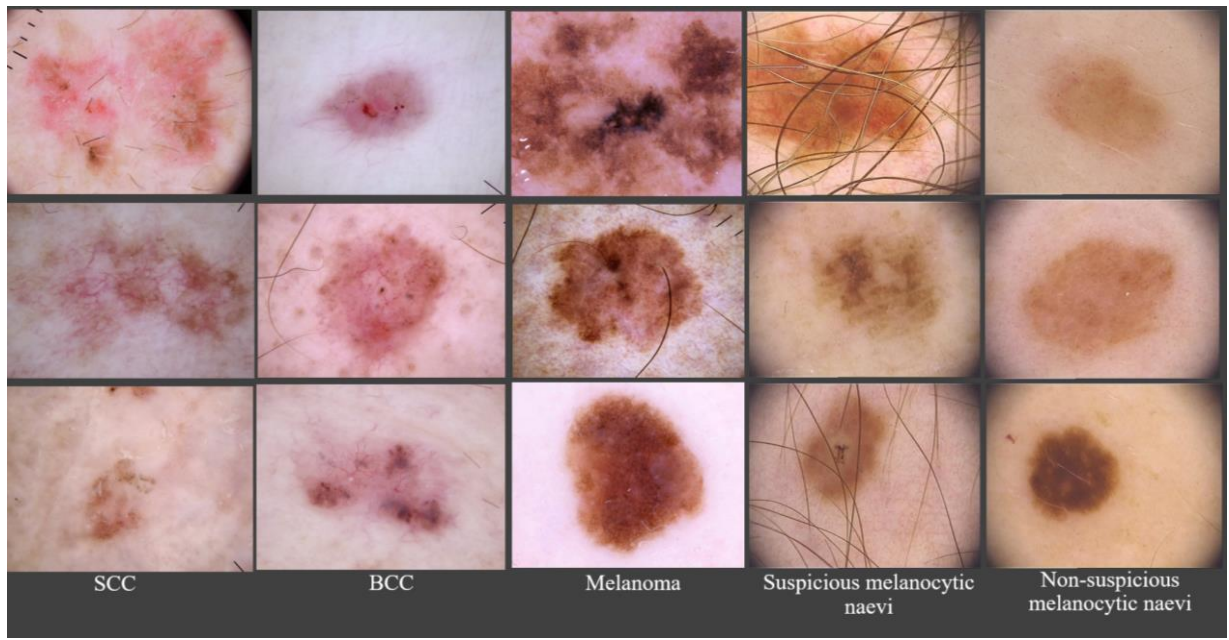

Figure 1: Examples of lesions belonging to different skin cancer types, along with suspicious and non-suspicious naevi. The cancer lesions were collected from ISIC 2019 dataset [11], while the naevi images are from a private dataset provided by The University of Queensland Diamantina Institute, Dermatology Research Centre.

While some naevi may be present from birth, such as congenital melanocytic naevi, they typically appear on the skin during childhood and continue to increase in number until early adulthood [12]. Although the majority of naevi are harmless, certain types such as dysplastic naevi may serve as precursors to melanoma [13] . Dysplastic naevi or atypical naevi are characterized by their irregular shape and a wide size range between 5 and 15 mm [14]. These naevi deviate from the typical appearance and exhibit features such as uneven borders, variegated colours, and asymmetry [13].

The number of naevi on an individual's body is generally the strongest phenotypic predictor of melanoma [15]. Therefore, the study and classification of naevi play a vital role in the detection of melanoma [16, 17]. Furthermore, researchers have proposed that while naevi exhibit considerable morphological diversity, all naevi within an individual can be grouped into a limited number of dominant categories or partitions. Consequently, a naevus that does not fit into any of these partitions may be considered suspicious or labelled as an "ugly duckling (UD) naevus" [18, 19]. This concept is employed for melanoma monitoring, as an UD naevus stands out as unusual and could potentially signal the existence of a melanoma [20].

Despite receiving special training and using dermatoscopes, dermatologists' sensitivity in detecting melanoma rarely exceeds 80%, with much lower sensitivity observed by general practitioners [21]. As a result, there is a growing need to incorporate Machine learning (ML) and computer-aided diagnosis (CAD) systems in clinical practice to improve the accuracy of melanoma diagnosis.

This article offers a comprehensive examination of state-of-the-art ML and DL-based algorithms designed for detecting melanoma and identifying suspicious and UD naevi. While there have been reviews conducted on melanoma and skin cancer detection using ML methods [5, 22-24], this study, stands out as the first to focus on suspicious naevi and UD naevus by applying ML models. The article is structured in the following manner: Section 2 provides an overview of AI, ML, and CAD. In Section 3, the focus is on how ML can be used for detecting melanoma. The section covers comparative studies involving CNN and dermatologists. In Section 4, the article delves into UD naevus signs and studies that use ML to identify suspicious naevi and UD nevi. Finally, Section 5 presents the discussion segment.

## 2. An Overview of Artificial Intelligence and Machine Learning

### 2.1. Applications of AI and ML

AI represents a comprehensive concept comprises the use of computer systems to simulate intelligent behaviour while minimizing human intervention. AI has found widespread use in various applications within the healthcare industry, including the field of dermatology [25]. While AI has made significant contributions to the detection of skin cancer, its integration into dermatology practice falls behind that of radiology. However, with continuous advancements and increased adoption of AI technologies, its accessibility is expanding, even reaching the general population [25]. AI holds potential in aiding the early detection of skin cancer, for instance, through the utilization of deep convolutional neural networks capable of analysing skin images for diagnostic purposes [25].

Machine learning (ML) has significantly expanded in research, finding its application across various domains such as text mining, spam detection, video recommendation, image classification, and multimedia concept retrieval [26]. ML is a branch of AI (Figure 2) that utilizes statistical models and algorithms to iteratively learn from data, enabling the prediction of characteristics for new samples and the accomplishment of specific tasks [25].

Deep Learning (DL), a subset of ML, is influenced by the information processing mechanisms observed in the human brain. Unlike conventional Machine Learning techniques that depend on predetermined rules, DL employs vast amounts of data to establish correlations between input and associated labels. Artificial Neural Networks (ANNs) comprises multiple layers of algorithms that constitute DL. Each layer provides a distinct understanding of the input data [27]. Conventional ML approaches entail a step-by-step process comprising pre-processing, feature extraction, meticulous feature selection, learning, and classification. The effectiveness of these techniques is heavily dependent upon the employed feature selection procedure as an imbalanced selection may result in inaccurate class discrimination. Conversely, DL can facilitate automatic learning of multiple task-oriented feature sets, distinguishing it from traditional ML methodologies. The process of learning and classification has been revolutionized by DL, which allows both tasks to be completed simultaneously. This approach has greatly increased the popularity of DL as a machine learning algorithm, particularly due to its effective use in processing big data [28]. As DL continues to evolve, it consistently provides new performance improvements across different machine learning tasks [29, 30]. Additionally, it simplifies progress in various learning areas [31] such as image super-resolution [32], object detection [22, 33] and image recognition [34]. Notably, DL has recently outperformed human ability in tasks like image classification.

Additionally, DL techniques have exhibited impressive aptitudes in replicating human abilities in a wide range of domains, such as medical imaging [27]. Within radiology, the main goal is often to identify structural irregularities and group them into specific disease classifications. Many Machine Learning algorithms with distinct implementations, mathematical principles, and logical foundations have been utilized for these classification tasks over time [35]. Consequently, multiple CAD systems were created and incorporated into clinical routines as early as the 2000s [35]. Convolutional neural networks (CNNs) represent deep neural networks meticulously structured for image analysis. They are typically trained through supervised learning, involving labelled data like dermoscopic images accompanied by their respective diagnoses or ground truths. This training enables CNNs to establish associations between input data and labels, empowering them to employ learned operations on unfamiliar images and classify them according to extracted features [5]. For a more in-depth understanding of CNNs, refer to the study by Chandra et al. [36], where they have discussed different kinds of CNNs in their Background section.

DL has various applications in the medical field, particularly in identifying and diagnosing diseases. For example, DL models have been utilized to diagnose conditions such as diabetes [37], breast cancer [38], chronic diseases [39], Alzheimer's disease [40] and more. Notably, Google AI [41] achieved an average accuracy of 70% in detecting prostate cancer, surpassing the performance of US board-certified general pathologists who achieved 61%. In 2020, DL played a vital part in diagnosing COVID-19 and is now widely used as a primary tool for automatic classification and detection of COVID-19 using chest X-ray images and other medical images in numerous hospitals worldwide [29]. Lastly, Esteva et al.[1] discovered that a DL network exhibited comparable diagnostic capabilities to twenty-one board-certified dermatologists when analysing a vast dataset of 129,450 images encompassing 2032 diseases.

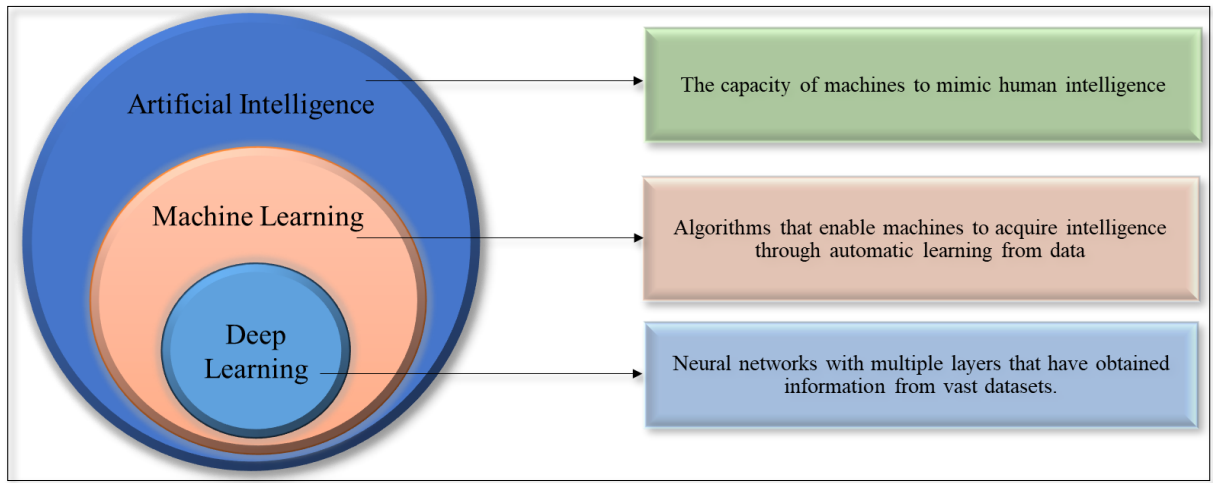

Figure 2: A visual representation displaying the relationship among AI, ML, and DL.

## 2.2. Computer Aided Diagnosis CAD

CAD has gained significant interest and has emerged as a crucial area of research in recent times [42]. This field utilizes machine learning methods to examine diverse forms of patient data, including both images and non-image data. The main objective is to evaluate the medical condition of the patient quantitatively. The assessments generated by CAD aim to aid healthcare professionals in their decision-making process by offering valuable insights and supporting clinical judgment. From a technical standpoint, CAD systems are computational tools that harness the power of machine learning algorithms to assist clinicians in interpreting complex medical data, ultimately enhancing diagnostic accuracy beyond what could have been achieved otherwise [42].

CAD has demonstrated its effectiveness in various sectors of healthcare, with dermatology being a notable field of application. Particularly in dermatology, CAD systems have been developed and employed to aid healthcare professionals in diagnosing skin conditions, including melanoma. These systems utilize advanced algorithms and machine learning techniques like Deep (CNNs) to analyse dermoscopic images and other relevant patient data.

CNN is a widely recognized and commonly utilized algorithm [43]. One of the advantages of CNN over previous algorithms is its ability to automatically identify important features without human supervision [44]. The use of CNNs has been extensive in various fields including speech processing, computer vision, and face recognition [45-47] . The design of CNNs is inspired by neurons found in animals and humans, specifically the visual cortex in cats. As a result of this similarity to biological neural networks, CNNs are able to simulate complex sequences of cells [44].

CNNs have garnered significant attention as a machine learning technique that holds promise in the field of melanoma prediction. The utilization of CNNs enables the analysis of extensive datasets, offering the potential to improve the accuracy and efficiency of melanoma diagnosis. As ongoing research and development efforts persist in this domain, it becomes evident that computer vision will assume a progressively significant role in the future of healthcare [1, 36, 48-50].

## 3. Exploring the Role of Machine Learning in Melanoma Detection.

### 3.1. Skin Lesion Classification/Detection

The use of Deep CNN techniques has revolutionized image classification, but applying these techniques to medical image analysis has been hampered by the scarcity of labelled data. The study by Pham et al. [48] offered a solution to this issue by proposing a classification model that incorporates Data Augmentation to improve the performance of skin lesion classification using Deep CNN. The study evaluated the proposed classification system on a dataset of 6,162 training images and 600 testing images of skin lesions, achieving state-of-the-art performance with higher AUC, AP, and ACC scores than traditional methods. Additionally, the study examined the influence of each image augmentation on the three classifiers and found that each augmentation affects the performance of each classifier differently. The study's findings suggested that applying data augmentation can help generate more samples and improve the accuracy and reliability of diagnoses in skin cancer classification and other medical image classification tasks. This study has provided a promising avenue for the use of Deep CNN in medical image analysis, even with limited labelled data.

In the pursuit of better melanoma detection, the authors of a recent study by Jojoa Acosta et al. [3] proposed a two-stage classification method. The first stage involved using Mask R-CNN to create a bounding box around the skin lesion, while minimizing visual noise. The output of this stage was a cropped bounding box that contained the skin lesion. In the second stage, the authors used ResNet152 to classify the cropped lesions. While Mask R-CNN can also be used for image classification, the authors chose to separate the two tasks in order to use higher-performing classification models. The proposed model was evaluated using the database associated with the challenge set out at the 2017 International Symposium on Biomedical Imaging ISBI. This approach allowed the authors to achieve high accuracy in melanoma detection, outperforming other state-of-the-art methods. By utilizing both object detection and classification models, the authors were able to reduce the impact of visual noise and improve the accuracy of their results. Overall, this two-stage classification method represents a promising approach to improving melanoma detection, and it could have significant implications for the early detection and treatment of skin cancer.

There is a study introduced by Alwakid et al. [51] proposed a Deep Learning (DL) approach for accurately extracting a lesion zone from skin lesion images. The proposed method consists of several steps, including image enhancement using Enhanced Super-Resolution Generative Adversarial Networks (ESRGAN) to improve image quality, segmentation to extract Regions of Interest (ROI) from the full image, data augmentation to rectify data disparity, and classification using a convolutional neural network (CNN) and a modified version of Resnet-50. To evaluate the proposed method, the authors used an unequal sample of seven types of skin cancer from the HAM10000 dataset. The results showed that the proposed CNN-based model outperformed the earlier study's results by a significant margin, achieving an accuracy of 0.86, a precision of 0.84, a recall of 0.86, and an F-score of 0.86.

Another study conducted by Li and Shen aimed to detect melanoma early by implementing three tasks: lesion segmentation, lesion dermoscopic feature extraction, and lesion classification [52]. The authors proposed two main networks, Lesion Indexing Network (LIN) and Lesion Feature Network (LFN), to achieve these tasks. The LIN consisted of two very deep fully convolutional residual networks (FCRN-88) and a lesion indexing unit (LICU). The FCRNs were trained using various training sets to produce segmentation and coarse classification. The LICU enhanced the coarse skin lesion possibilities maps from the FCRNs by employing distance maps that measure the importance of each pixel to refine the coarse lesion classification. The LFN was designed to deal with the task of dermoscopic feature extraction. It is a convolutional neural network (CNN) that was trained using patches extracted from dermoscopic images. The LIN method was compared with other models, including the FCN designed by Long et al [53], the Unet [54], the fully convolutional Inception (II-FCN), and the encoder-decoder network using RNN layer (Auto-ED). The LIN method achieved the best outcome compared to the results of the other models. The LFN was able to gain the best average precision and sensitivity score for dermoscopic feature extraction, which proved the efficiency of this design to complete the challenge successfully. Overall, the study by Li and Shen showcased the effectiveness of using deep learning networks, such as LIN and LFN, for early detection of melanoma. The results suggest that these networks could potentially be used in clinical settings to aid in the early detection of melanoma and improve patient outcomes.

A study by Yu et al. [55] was carried out to identify melanoma at an early stage. In order to extract more useful features and achieve better recognition accuracy, researchers used very deep networks with over 50 layers. A fully convolutional residual network (FCRN) was constructed to accurately segment skin lesions, and a multi-scale contextual information integration scheme was incorporated to enhance its capability. To build a two-stage network structure for segmentation and classification, the FCRN for segmentation was combined with very deep residual networks for classification. For the classification process, the same structure as FCRN was used, along with an additional 7*7 average pooling layer followed by 16 residual blocks, a softmax classifier and a support vector machine (SVM) classifier. The final classification output was derived from the average of predictions acquired from these classifiers. The proposed approach was evaluated using the ISBI 2016 Skin Lesion Analysis Towards Melanoma Detection Challenge dataset and it achieved the best results among others with 50 layers in both classification and segmentation categories. The study demonstrated that using very deep networks is effective in early detection of melanoma and combining segmentation and classification in a two-stage network structure is a successful approach.

Furthermore, a study by Kumar et al. [56] aimed to classify benign and malignant skin lesions using a deep learning and Support Vector Machine (SVM) approach. They utilized a dataset comprising 2637 training images from the International Skin Image Collaboration (ISIC) archive, along with 660 images for testing. The PH2 dataset was also included in the study. The research consisted of three main stages: lesion segmentation, feature extraction, and classification. For lesion segmentation, they employed the U-Net architecture, specifically designed for biomedical image segmentation, which effectively delineates lesion boundaries using convolutional neural networks. Additionally, transfer learning techniques were applied using five pre-trained convolutional neural network models (AlexNet, VGG16, ResNet-50, InceptionV3, and DenseNet201) to extract relevant features from the segmented lesions. These models are known for their exceptional performance in image classification and were used for feature extraction and subsequent classification, allowing for a comparative analysis of their effectiveness in skin lesion classification. In the final stage, the SVM algorithm was employed for classification. The SVM was trained on the extracted features to distinguish between benign and malignant skin lesions. Among the investigated models, DenseNet201 exhibited the highest accuracy of 89% when combined with SVM, demonstrating its effectiveness as the most suitable model for this task.

In addition, a study carried out by Ercument and Maria aimed to classify malignant and benign skin lesions using various deep learning models, specifically AlexNet, GoogLeNet, and ResNet50 CNNs [57]. Dermoscopic images were utilized for the research, incorporating a dataset consisting of 19,373 benign and 2,197 malignant lesions obtained from the ISIC database. All three CNNs, renowned as previous winners of the ImageNet competition, were reconfigured for binary classification purposes. The dataset was divided into training (80%) and validation (20%) sets. Uniform parameters were employed across all experiments for each CNN model. The results revealed that the ResNet50 model achieved the highest performance with a classification accuracy of 92.81%. Moreover, AlexNet demonstrated superior time complexity in comparison to the other models.

In the study conducted by Kasmi and Mokrani [58] the researchers implemented an automatic ABCD (asymmetry, border irregularity, colour, and dermoscopic structure) scoring system to distinguish between malignant and benign skin lesions. The preprocessing phase involved the application of a median filter to remove artefacts such as bubbles and thin hair. This step facilitated the automatic detection of hair using Gabor filters. Lesions were subsequently segmented using a geodesic active contour (GAC) approach. Algorithms were developed to extract the characteristic features associated with the ABCD attributes. Multiple methods were employed to detect colour asymmetry and dermoscopic structures. To classify the lesions, the total dermoscopic score (TDS) was computed. A low TDS indicated a benign lesion (TDS < 4.75), a middle TDS was interpreted as a suspicious case (4.75 < TDS < 5.45), and a high TDS indicated a higher likelihood of a malignant melanoma (TDS > 5.45). The experimental evaluation involved the use of 200 dermoscopic images, comprising 80 malignant melanomas and 120 benign lesions. The algorithm achieved a sensitivity of 91.25% and specificity of 95.83%, comparable to the reported 92.8% sensitivity and 90.3% specificity achieved by human implementation of the ABCD rule. Furthermore, the method demonstrated an overall accuracy of 94.0%.

Furthermore, a study by F. Dalila et al., [59] aimed to differentiate between malignant and benign skin lesions through a comprehensive three-stage process involving skin lesion segmentation, feature extraction, and classification. In this research, an Ant colony-based segmentation algorithm was employed, considering three distinct types of features to describe malignant lesions, namely geometrical properties, texture, and relative colours. Relevant features were selected from these categories, and two classifiers, namely K-Nearest Neighbour (KNN) and Artificial Neural Network (ANN), were utilized for classification purposes. To achieve the objectives of the study, a dataset comprising 172 dermoscopic images was employed, encompassing 88 malignant melanomas and 84 benign lesions. The findings demonstrated that a subset of 12 features (out of 112 extracted) proved sufficient for accurately detecting malignant melanoma. Furthermore, the K-Nearest Neighbours classifier achieved a correct classification rate of 85.22% for the tested images when automatic masks were applied, compared to 87.50% with manual masks. In contrast, the Neural Network demonstrated a higher level of success in correctly classifying the tested images, achieving an accuracy of 93.60% with automatic masks compared to 86.60% with manual masks.

While there are many methods available to identify melanoma from benign skin conditions using dermoscopic images, there are limited non-invasive methods available identify the stage or type of melanoma cancer. Thus, there is a study that introduced two classification systems that used the same algorithm based on the thickness of melanoma skin cancer to identify its stage or type [4]. The first proposed system classifies melanoma into two categories based on tumour thickness - tumours with thickness less than 0.76 mm are classified as the first stage, while tumours with thickness greater than or equal to 0.76 mm are classified as the second stage. The second system classifies melanoma into three categories or stages based on tumour thickness - the first stage has a tumor thickness less than 0.76 mm, the second stage has a tumour thickness between 0.76 mm and less than 1.5 mm, and the third stage has a tumor thickness greater than 1.5 mm. to run theses system, Convolutional Neural Networks (CNN) was used along with Similarity Measure for Text Processing (SMTP) as loss function.

TABLE I. Summary of the studies covered in this section, encompassing the employed ML models, datasets, objectives, and outcomes.

| Study | Methodology | Dataset | The study aim | Results |
|---|---|---|---|---|
| **Pahm et al. [48]** | Inception V4 for feature extraction as well as three type of classifiers including random forest, support vector machine and neural network (NN). data augmentation has been applied as well. | ISIC 2017 dataset (ISBI Challenge [60]) | Differentiate melanoma from non-melanoma. | The data augmentation and NN technique achieved state-of-the-art performance in classifying melanoma, with higher AUC, AP, and ACC scores than the previous best (89.2% vs 87.4%, 73.9% vs 71.5%, and 89.0% vs 87.2%). |
| **Jojoa Acosta et al. [3]** | Method involves using Mask and Region-based CNN to automatically crop the region of interest within a dermoscopic image and ResNet152 structure for lesion. Data augmentation | ISIC 2017 dataset (ISBI Challenge [60]) | Differentiate melanoma from non-melanoma. | The overall accuracy of the model was 0.904, with a sensitivity of 0.820 and a specificity of 0.925, surpassing other state-of-the-art |

| | technique was applied as well. | | | methods in terms of performance. |
|---|---|---|---|---|
| **Alwakid et al. [51]** | The image is improved using ESRGAN and segmentation to isolate the ROI. Data augmentation addresses data imbalance. Skin lesion classification is done by analysing the image with a CNN and adapted Resnet-50. | HAM10000 | Diagnosing seven different types of cancer. | The research outperformed alternative approaches with a precision score of 0.84, accuracy score of 0.86, recall score of 0.86, and an F-score of 0.86. |
| **Li and Shen [52]** | A new deep learning structure that consists of two fully convolutional residual networks (FCRNs) is presented to simultaneously produce the segmentation and initial classification outcomes. The results of the initial classification are improved by using a lesion index calculation unit (LICU)that calculates a distance heat-map to refine the outcome. | ISIC 2017 | The study aimed to achieve early detection of melanoma by implementing three tasks: lesion segmentation, dermoscopic feature extraction from lesions, and lesion classification. | The framework has shown promising accuracy in the experiments conducted, achieving 0.753 for task 1, 0.848 for task 2, and 0.912 for task 3. |
| **Yu et al. [55]** | A convolutional residual network with multiple-scale feature representations is created at first to segment skin lesions thoroughly. Then, a similarly deep residual network is employed to distinguish melanomas from non-melanoma lesions based on the segmentation results. | ISBI 2016 Skin Lesion Analysis Towards Melanoma Detection Challenge dataset. | Classify the skin lesion as either melanoma or non-melanoma | The results of the experiment demonstrate notable progress made by this system, establishing it as the top performer in classification and ranking second in segmentation out of 25 and 28 teams, respectively. |
| **Kumar et al [56]** | The study comprised three primary stages: lesion segmentation, feature extraction, and lesion classification. The U-Net method was used for segmentation. For feature extraction, AlexNet, VGG16, ResNet-50, InceptionV3, and DenseNet201 were employed, and SVM was utilized for lesion classification. | Dataset comprising 2637 training images from the ISIC archive, along with 660 images for testing. The PH2 dataset was also used. | Classify benign and malignant skin lesions | In this study, the combination of SVM and DenseNet201 resulted in an 89% accuracy, which was the highest achieved. |
| **Ercument and Maria [57]** | This study investigated the effectiveness of AlexNet, GoogLeNet, and ResNet50 convolutional neural networks (CNNs) in | 19,373 benign and 2,197 malignant dermoscopic lesions were obtained from ISIC for analysis. | Classify the lesion as either melanoma or | The ResNet50 model exhibited the best performance in the experiments, achieving a |

| | classifying dermoscopic images | | non-melanoma. | classification accuracy of 92.81%. In terms of time complexity measurements, AlexNet was ranked as the most efficient. |
|---|---|---|---|---|
| **Kasmi and Mokrani [58]** | The study used an ABCD scoring system. A median filter was applied in the preprocessing phase to remove artefacts, followed by hair detection with Gabor filters. Lesions were segmented using a geodesic active contour approach, and algorithms were developed to extract characteristic features related to ABCD attributes. The TDS was computed for lesion classification. | Study used 200 dermoscopic images - 80 malignant melanomas and 120 benign lesions. | Classify the lesion as melanoma or benign naevus | Algorithm sensitivity is 91.25% and specificity is 95.83%, comparable to human ABCD rule implementation with sensitivity of 92.8% and specificity of 90.3%. |
| **Dalila et al. [59]** | Lesion segmentation is done using Ant Colony Optimization. Three types of features are then extracted: geometrical properties based on the ABCD rule, texture features and relative colour. A total of 112 features are extracted, 12 relevant attributes selected using Relief algorithm. For classification, K Nearest Neighbour (KNN) and Neural Network (NN) classifiers are used. | 172 dermoscopic images tested, 88 malignant melanomas and 84 benign lesions. | Distinguish between malignant melanoma and benign lesions | The KNN and manual masks had 85.22% and 87.50% identification rates, respectively. The NN outperformed with 93.60% accuracy using automatic masks, compared to the 86.60% achieved with manual masks. |
| **Patil and Bellary [4]** | The study was conducted modified CNN, employing various loss functions, including the Similarity Measure for Text Processing (SMTP). | The dataset, available at https://www.uco.es/grupos/ayrna/ieeetmi2015, has 81 features and is divided into binary and multi-class datasets. It includes 250 images of melanoma cancer categorized by size: 167 < 0.76 mm, 54 between 0.76 and 1.5 mm, and 29 > 1.5 mm. | Categorize the different stages of melanoma cancer. | This approach achieved better accuracy, specificity, and sensitivity than SVM and normal CNN. |

### 3.2. Comparative Studies: CNN vs. Dermatologists

In a ground-breaking study introduced by Esteva et al. [1], researchers have presented a noteworthy investigation that showcases the efficacy of a singular CNN in the classification of skin lesions, utilizing exclusively pixel data and disease labels as inputs. The CNN was subjected to an end-to-end training process using an extensive dataset comprising 129,450 clinical images, surpassing previous datasets by two orders of magnitude in scale. To evaluate the CNN's performance, the researchers conducted comprehensive assessments involving 21 board-certified dermatologists, employing biopsy-proven clinical images in two crucial binary classification scenarios: distinguishing keratinocyte carcinomas from benign seborrheic keratoses and identifying malignant melanomas versus benign naevi. The former case pertains to the identification of the most prevalent cancers, while the latter pertains to the identification of the most fatal skin cancer. Remarkably, the CNN demonstrated performance comparable to all experts tested across both tasks, underscoring its capability as an artificial intelligence system proficient in skin cancer classification, at a level on par with dermatologists. This groundbreaking achievement carries profound implications for the future of dermatology, as the implementation of deep neural networks on mobile devices holds the potential to expand the reach of dermatologists beyond traditional clinical settings.

Another study conducted by P. Tschand et al., [49] aimed to differentiate between benign and malignant pigmented skin lesions. The study involved a comprehensive analysis by training a CNN-based classification model to evaluate two imaging methods. The CNN models employed in this experiment included two distinct architectures, namely InceptionV3 and ResNet50. The dataset used for training and testing consisted of a total of 7,895 dermoscopic images and 5,829 close-up images of lesions obtained from a primary skin cancer clinic. The dataset encompassed cases recorded between January 1, 2008, and July 13, 2017. The performance of the combined CNN model was assessed on an independent sample set comprising 2,072 cases, and the results were compared to the evaluations of medical personnel serving as human raters. Ninety-five clinicians acted as human raters for the study, including 62 board-certified dermatologists with varying levels of experience in dermoscopy. Based on their experience levels with dermoscopy, they were placed into three groups: beginner raters (<3 years), intermediate raters (3-10 years), and expert raters (>10 years). Evaluation using the area under the receiver operating characteristic curve revealed that the combined CNN model demonstrated superior overall diagnostic capability compared to human raters, achieving a higher sensitivity (80.5%; 95% CI: 79.0%-82.1%) when specificity was fixed at the mean level observed among human raters (51.3%). However, there was no significant difference between the performance of the combined CNN and expert raters in terms of specific diagnostic accuracy (37.3%; CI: 35.7%-38.8% vs. 40%; CI: 37%-43%, respectively). These findings provide empirical evidence that CNNs can attain a level of performance comparable to that of humans in distinguishing benign and malignant lesions in both binary classification and multiclass tasks.

A significant study that aimed to utilize Google's Inception v4 architecture in a deep learning CNN to train, validate, and test dermoscopic images of melanocytic lesions for diagnostic classification (i.e. melanoma and benign naevi)[61]. The researchers sought to evaluate the performance of the CNN against a large cohort of dermatologists with varying levels of expertise. A comparative cross-sectional reader study was conducted using a 100-image test-set, where the dermatologists assessed the images at two levels: level-I involved dermoscopy alone, whereas level-II included supplementary clinical information and images. The principal measures that were evaluated include the sensitivity, specificity, and the area under the curve (AUC) of the receiver operating characteristics (ROC) to classify lesions through CNN as compared to dermatologists for diagnostic purposes. The secondary objectives of this study comprise assessing the dermatologists' performance in diagnosing lesions and determining any dissimilarities in their diagnostic proficiency between level-I and level-II of the reader study. At level-I, a team of dermatologists attained an average sensitivity of 86.6% and specificity of 71.3% for the classification of lesions. Subsequently, the integration of clinical data in level-II enhanced sensitivity to 88.9% and specificity to 75.7%. The ROC curve generated by the CNN demonstrated greater specificity (82.5%) than that achieved by dermatologists at level-I (71.3%) and level-II (75.7%), with comparable sensitivities. Moreover, the AUC was higher than the mean AUC calculated for dermatologists (0.86 versus 0.79). Further, the CNN's efficacy was assessed in contrast to the leading five algorithms from the 2016 International Symposium on Biomedical Imaging (ISBI) challenge, and it yielded outcomes congruent to those of the top three algorithms. The CNN surpassed most dermatologists concerning diagnostic precision. Consequently, these results indicate that dermatologists, regardless of their expertise, could leverage a CNN's aid for tasks relating to image classification.

Lastly, study has addressed the imbalanced dataset problem in melanoma prediction by proposing an appropriate CNN architecture that involves custom loss functions, mini-batch logic, and reformed fully connected layers [62]. The approach was tested on a training dataset consisting of 17,302 images of melanoma and naevus, which is the largest dataset used in melanoma prediction to date. The model was compared to the performance of 157 dermatologists from 12 university hospitals in Germany, all of whom used the same dataset for their evaluation. The results showed that the proposed approach outperformed all 157 dermatologists and achieved higher performance than the state-of-the-art approach, with an area under the curve of 94.4%, sensitivity of 85.0%, and specificity of 95.0%. Furthermore, using the best threshold provided the most balanced measure compared to other studies, with a sensitivity of 90.0% and specificity of 93.8%. These results indicate that the proposed approach has significant potential for medical diagnosis. Overall, this study highlights the potential of computer vision and AI-based solutions for enhancing the accuracy and efficiency of melanoma diagnosis, which could ultimately lead to improved patient outcomes.

TABLE II.   Summary of the studies covered in this section, including the employed ML models, datasets, objectives, and outcomes.

| Study | Methodology | Dataset | The study aim | Results |
|---|---|---|---|---|
| **Tschand et al.[49]** | The study used two CNN models, InceptionV3 and ResNet50, and 95 medical staff acted as raters. | The study used two imaging systems. The training set has 7,895 dermoscopic images and 5,829 close-up images. The validation set has 340 dermoscopic images and 635 close-up images. | Distinguish between benign and malignant pigmented skin. | Results showed that the combined CNN model was better than human raters at overall diagnosis but performed similarly to expert raters in specific diagnoses accuracy |
| **Haenssle et al. [61]** | A study was implemented using Google's Inception v4 architecture. The CNN was compared to dermatologists with varying levels of expertise in a reader study. | The study was conducted using a 100-image test-set, collected from ISIC 2016 (ISBI), where the dermatologists assessed the images at two levels: level-I involved dermoscopy alone, whereas level-II included supplementary clinical information and images | Diagnose melanocytic lesions (melanoma vs benign lesion) | The CNN outperformed most dermatologists in diagnostic precision, even when compared to leading algorithms from an international challenge |
| **Pham et al. [62]** | The study was implemented using Optimized deep-CNN architecture with custom mini-batch logic and loss function used in approach. | The performance evaluation is conducted using the ISIC 2019 challenge dataset, which includes 17,302 images of melanoma and naevus. | Melanoma diagnosis and addressing the issue of imbalanced datasets in medical imaging. | The study outperformed all 157 dermatologists and achieved higher performance than the state-of-the-art approach, with an AUC of 94.4%, sensitivity of 85.0%, and specificity of 95.0% |
| **Esteva et al. [1]** | GoogleNet Inception v3 CNN architecture was used to carry out this study | The study was tested on set of 129,450 clinical images including 3,374 dermoscopy images | The study aimed to implement two crucial binary classification: distinguishing keratinocyte carcinomas from benign seborrheic keratoses and | The CNN demonstrated performance comparable to all experts tested across both tasks. |

| | | | identifying malignant melanomas versus benign naevi. | |

# 4. Detection of Ugly duckling lesions

## 4.1. The 'Ugly Duckling' Sign in Cutaneous Lesions

The standard method for identifying malignant melanoma through visual examination is known as the ABCDE (Asymmetry, Border, Colour, Diameter, Evolution) criteria [63]. This criterion indicates that a melanoma is said to be a spot that lacks in symmetry (Asymmetry), has a spreading or an irregular edge (Border), a variegated colour (Colour), increasing in diameter and getting bigger (Diameter), and changing its size and colour over time (Evolution). Nonetheless, there are instances where malignant melanomas do not adhere to this ABCDE criteria, requiring an alternative approach for recognizing a malignant melanoma visually. Therefore, the concept of an 'Ugly Duckling' (UD) was introduced as an additional factor for observation of melanoma [20]. An ugly duckling is defined as a naevus that deviates in its characteristics from that of other surrounding naevi, considering that naevi in an individual resemble one another. Hence, ugly duckling naevi are considered suspicious for a possible melanoma.

## 4.2. AI-Based Methods for Ugly Duckling and Suspicious Naevi Identification

While dermatology has seen the advantages of recent developments in deep learning techniques applied in medical image domains, only few studies have explored the use of deep learning methods for the identification of the ugly duckling sign. This could be because currently there are no public datasets that are specifically annotated with ugly duckling labels. Most widely used skin lesion datasets such as ISIC, HAM10000, and PH2 include melanoma and benign labels for lesions [64]. In addition, with the exception of the ISIC 2020 dataset, most datasets have focused on collecting single lesion images, whereas detecting the ugly duckling sign requires multiple images per person to incorporate the within person contextual information. However, not all melanoma lesions might display the ugly duckling sign, and melanoma characteristics cannot be used interchangeably with the ugly duckling sign [65]. Hence, studies focusing on identification of ugly ducklings would either have to collect their own data, and specifically label the datasets for ugly duckling lesions.

Among the few works that have focused on the ugly duckling sign, the use of wide-field images from total body photography are popular [66-70]. This is an advantage in ugly duckling related work as by definition, an ugly duckling is a contextual observation in comparison to neighbouring lesions on a body-site, and wide-field images provide a way to include this contextual information.

The earliest work focusing on wide-field images for ugly-duckling identification by Birkenfeld et al. [66] uses a logistic regression to build a computer-aided classification system. For a population of 133 patients, they crop single lesions with a diameter > 3mm from the wide-field images of different body-sites. These lesion images are manually labelled as suspicious or non-suspicious by a board-certified dermatologist based on 399 features included within the ABCD criteria. An optimized $L_2$ penalized logistic regression model with class balancing is then trained as a classifier for the cropped lesions, resulting in 0.89 AUC, 84% sensitivity for suspicious lesions and 72.1% specificity for non-suspicious lesions. They identify that a threshold value of 0.46 achieved the lowest false positive rate in the classifier while still adhering to their clinical design criteria of 95% true positive rate. Hence, they use this threshold score of 0.46 to make the predictions on the test set of lesions. All lesions below the threshold value are considered non-suspicious while all lesions with a value above the threshold are considered suspicious. With their contributions, they establish a suspiciousness score capable of distinguishing suspicious lesions from non-suspicious lesions that is aligned with naked-eye examinations. However, the advantage of using wide-field images is somewhat lost in their work as all cropped lesions are pooled together to train the classifier, resulting in the binary decision being more at lesion-level than patient-level.

A few works also employ self-supervised methods for analysing wide-field images in order to identify ugly ducklings. Mohseni et al. [67] proposes an outlier detection approach where ugly ducklings deserving of more attention from examining physicians are identified. They first detect all lesions from wide-field images using a Single Shot Detector (SSD) network, which are then extracted by segmentation with a U-Net variant. Ugly duckling lesions in these images were labelled by a board-certified dermatologist. For the collection of lesions from each wide-field image, outlier detection is then performed by training a Variational Auto Encoder. They also create a ranking system for the lesions by generating an embedding for each lesion and calculating the L2 distance between them. A threshold for lesions from each wide-field image is calculated to decide whether a lesion is an ugly duckling or not. Their system achieves a sensitivity of ~72%, specificity of ~94%, and an overall accuracy of ~94%. The study by Useini et al. [69] proposes a similar method to the above by employing a self-supervised outlier detection to identify ugly duckling lesions. They first use a YOLOR model to detect and extract lesions from wide-field images of dorsal regions of patients. A DINO model is then used for self-supervised learning, which identifies a distance threshold between lesions which is proposed as an ugly duckling score. These studies demonstrate that patient-level analysis of wide-field images is more advantageous for identifying ugly ducklings accurately.

More recently, Garcia et al. [70] evaluated the use of unsupervised autoencoders to detect suspicious skin lesions in the form of outlier detection. They manually extract around 1800 lesion images from wide-field images, which are labelled as benign or malignant. The autoencoder outlier detector is trained with 90% of the benign lesions and evaluated on a test set containing 180 benign lesions and 20 hand-selected melanomas. Out of these 20 melanomas, 5 were melanomas with features that are not clearly distinguishable and more challenging to identify. However, 15 were clearly identifiable melanomas, which were used to define the accepted reconstruction error in the autoencoder, thereby defining the threshold for outlier detection. From the evaluation, the trained autoencoder was able to identify 2 of the 5 challenging melanoma cases correctly, failing to identify the remaining 3 due to no presence of obvious distinctive features. While the use of autoencoders as an efficient method for suspicious lesion detection is proven in this study, the patient-level contextual information is lost as they pool the extracted lesions from wide-field images to train the autoencoder. Furthermore, defining the outlier detection threshold based only on clearly presenting melanomas might also have affected the poor performance in identifying the more challenging melanomas as it introduces a bias. In addition, using specific melanoma and benign labels cannot be generalized to detect ugly ducklings as not all melanomas present ugly duckling characteristics. The study may have benefitted more by using specific ugly duckling labels for the dataset.

Apart from using wide-field images, to our knowledge only one known work exists in current literature for analysing individual lesion images to identify ugly ducklings. Yu et al. [71] uses the dataset from SIIM-ISIC2020 melanoma classification challenge containing about 30,000 individual lesion images from around 2000 patients, labelled as melanoma or benign. This study recognizes that analysing images at only lesion-level is not useful for ugly duckling identification and takes important steps to incorporate more contextual information. First, they extract multi-scale features for all lesions per-patient using a deep neural network. Then, these features are used to learn a patient-specific contextual embedding by modelling the dependencies among lesions using a Transformer encoder. These contextual embeddings are then used to perform patient-level and lesion-level predictions concurrently. To optimize the predictions and address the data imbalance issue, they also propose a group contrastive learning strategy to rank lesions of an individual based on their appearance similarity. The optimized model achieves ~90% AUC with ~82% for both sensitivity and specificity. However, although patient-level contextual information is included in the training for this model, the ground truth annotations are again not specifically ugly duckling, and as mentioned earlier the definition of an ugly duckling can not be equated to that of melanoma.

Furthermore, a research team from the United States implemented the "ugly duckling naevi" concept to differentiate between suspicious and non-suspicious pigmented lesions [68]. Their approach involved the utilization of deep convolutional neural networks on a dataset comprising 38,283 lesions from 135 wide-field clinical images collected from 133 patients at the Hospital Gregorio Maranon in Madrid, Spain. The algorithm was applied to participant images on a wide-field scale, followed by blob detection and subsequent cropping of the detected blobs. The cropped lesions were then inputted into the classification model. The deep convolutional network was employed to identify suspicious lesions based on labels provided by dermatologists. Furthermore, the network aided in the detection of "ugly duckling" naevi by extracting deep features and employing heat map techniques for visualization. The model achieved a sensitivity of over 90.3% and specificity of 89.9% in distinguishing suspicious lesions from non-suspicious lesions, as well as from normal skin and complex backgrounds. Additionally, the researchers defined the criteria for identifying "ugly duckling" naevi as the patient-dependent probability of each lesion being suspicious relative to the other observable lesions within the wide-field image. This definition allowed them to calculate an "ugly duckling" score using the cosine distance between the output feature vector of a lesion from the deep convolutional neural network and the averaged geometric feature centre of all observable lesions. These scores were then utilized to create a ranking system for the lesions, which was compared with the rankings performed by three board-certified dermatologists to assess agreement. The outcomes of this study provide clear evidence that well-optimized deep learning methods can be effectively employed to accurately assess the suspiciousness of pigmented lesions [68].

Lastly, In our previous study [72], we aimed to investigate the common visual characteristics shared by both suspicious and non-suspicious naevi, with the goal of enhancing the identification and understanding of key visible properties that enable their classification into their respective groups. The study comprised two main stages: the extraction of naevi features and the subsequent classification of naevi into suspicious and non-suspicious categories. Various machine learning (ML) methods were employed, including principal component analysis (PCA) and convolutional autoencoder (CAE), for feature extraction, followed by the implementation of random forest (RF) and artificial neural network (ANN) algorithms for naevus classification. The dataset used to evaluate these models consisted of 33,368 dermoscopic skin lesion images from 59 study participants. The skin lesion images were taken at multiple time points and were labelled by their visit number. Each image was manually labelled by a certified melanographer, resulting in a total of 26,606 images of non-suspicious naevi and 1,616 images of suspicious naevi. The dataset was partitioned based on the patients, with 14 patients for testing and 45 patients for training. By utilizing the features extracted by the CAE, the ANN achieved impressive average accuracy, specificity, sensitivity, precision, and area under the curve (AUC) values of 95.62%, 91.24%, 100%, 91.95%, and 95.6%, respectively. Additionally, the RF analysis revealed that both PCA and CAE-based methods yielded an overall accuracy of 88.46%. Furthermore, the RF algorithm was employed to rank the features, aiding in the selection of the most important features that proved useful for naevus classification. If clinically validated, ML approaches have the potential to serve as efficient guides in the early detection of melanoma by accurately identifying suspicious naevi that require careful assessment by clinicians.

TABLE III. *Summary of the studies covered in this section, including the employed ML models, datasets, objectives, and outcomes.*

| Study | Methodology | Dataset | The study aim | Results |
|---|---|---|---|---|
| **Birkenfeld et al. [66]** | The process involved colour correction, segmentation, feature extraction with 399 features under six categories (asymmetry, border, colour, texture, pixel size and lesion area), followed by Analysis of Variance and principal component analysis to select features for machine learning. Logistic regression was used for classification. | Images of major body parts from 133 patients with skin lesions were taken, and 1759 pigmented lesions were categorized as suspicious or non-suspicious. | Ugly identification and melanoma detection through body examinations to identify suspicious pigmented lesions. | The classifier had a 100% sensitivity for confirmed suspicious lesions and 83.2% for unconfirmed ones. Non-suspicious lesions had a sensitivity of 72.1%, and overall accuracy was 75.9%. |
| **Mohseni et al. [67]** | Lesions are detected and segmented using SSD and U-Net in wide-field images. VAE identifies outliers through lesion embedding L2 distance. YOLOR detects lesions from dorsal images for outlier identification. A self-supervised DINO model assigns a | 32 images were obtained from the SD198 [73], and SD-260 datasets [74] along with 43 more images which were collected from clinics. | Identification of UD naevus | The system has 72% sensitivity, 94% specificity, and 94% overall accuracy. |
| | UD score using a lesion threshold. | | | |
| Useini et al. [69] | Outlier detection involves using YOLOR to extract lesions from dorsal region images. Self-supervised DINO model determines a threshold between lesions as an ugly duckling score. | 91 patients' data collected at USZ dermatology clinic. | Automated lesion detection and self-supervised identification of UD naevus. | The detection algorithm, which were tested on two new patient images, achieved an average recall of 93% with an IoU threshold of 0.5. |
| Garcia et al. [70] | Unsupervised autoencoders to detect suspicious skin lesions through outlier detection | The USZ Dermatology Clinic uses Foto Finder, 1800 lesion images, for 91 patients, were classified as either benign or malignant. | Identify suspicious lesions using real-world data from consultations at the USZ Dermatology Clinic. | The algorithm achieved high sensitivity (0.87) and specificity (0.83) in detecting clear melanomas with an AUC of 0.9 |
| Yu et al. [71] | The study was implemented using CNN-Transformer. | ISIC 2020, SIIM-ISIC melanoma classification. The dataset has around 30000 images from 2000 patients | Melanoma detection model using the ugly duckling method. | The approach achieved 90% AUC with ~82% sensitivity and specificity |
| Soenksen et al. [68] | The researchers employed deep CNN. | The dataset has 38,283 lesions from 135 wide-field clinical images collected from 133 patients at Hospital Gregorio Maranon in Madrid, Spain. | Introduce the "ugly duckling naevi" concept to distinguish between suspicious and non-suspicious | The model achieved over 90.3% sensitivity and 89.9% specificity |
| Al Zegair et al. [72] | The study was carried out using various ML methods like PCA and CAE which were used to extract naevi features, followed by RF and ANN algorithms for classifying into suspicious and non-suspicious categories. | The dataset had 33,368 skin lesion images from 59 participants. Each image was labelled by a melanographer resulting in 26,606 non-suspicious naevi images and 1,616 suspicious naevi images. | Identify visual characteristics of suspicious and non-suspicious naevi to improve their classification. | The ANN achieved high accuracy, specificity, sensitivity, precision, and AUC values of 95.62%, 91.24%, 100%, 91.95%, and 95.6% by using CAE features. |

## 5. Discussion

Melanoma originates in the melanocyte cells of the skin and has the potential to appear on any part of the body. It is categorized as the most lethal form of skin cancer due to its capability to spread rapidly through an individual's body via blood or lymphatic system, if not diagnosed early [75]. Hence, timely and accurate identification of melanoma can be pivotal in saving a person's life and result in a 5-year survival rate of 95% [6]

Melanocytic naevi, which are benign pigmented lesions, develop as a result of the proliferation of melanocytes. The existence of multiple naevi indicates an increased risk of melanoma. Additionally, as many as 30% of melanomas may originate directly from such naevi[9, 10].

The concept of an "Ugly Duckling" was added as an additional factor to assist in the diagnosis of melanoma. An ugly duckling refers to a mole that has unique characteristics different from its adjacent moles. As moles within an individual usually have similarities and share a similar pattern, an ugly duckling mole appears unusual and may suggest the existence of a cancerous melanoma [20].

The study of CAD has gained significant importance in the realm of diagnostic radiology and medical imaging [76]. CAD combines machine learning techniques with patient analysis, using either imaging alone or both imaging and clinical data to enhance accuracy and streamline decision-making, potentially mitigating the shortage of specialized professionals [76]. Automated systems could play a critical role in skin cancer diagnosis, particularly where specialist availability is limited. By using CAD, healthcare providers can realize life-saving benefits while reducing costs. Early detection of melanoma risk through skin assessment is essential for effective treatment and favourable patient outcomes. The use of ML and DL techniques have surged in skin cancer classification as they efficiently tackle challenges and yield satisfactory results.

CNNs on the other hand, are complex neural systems that are specifically designed for analysing images. The primary training method for these networks is supervised learning, which involves the use of labelled data, such as dermoscopic images and their corresponding diagnoses or ground truths. Through this approach, CNNs can establish connections between input data and labels, enabling them to categorize unfamiliar images based on detected features using the acquired knowledge [5].The utilization of convolutional neural networks (CNNs) holds promise in augmenting and expanding clinically meaningful databases, particularly in the fields of clinical dermatology and dermatopathology (not discussed in this review) where visual pattern recognition plays a crucial role in diagnosis [52].

The review covers 21 studies, utilizing 14 different machine learning and deep learning techniques to classify skin lesions and/or detect melanoma. Seven of these studies concentrated on identifying suspicious naevi or UD naevi. These investigations used various machine learning strategies to perform tasks like skin lesion segmentation, feature extraction, and classification. Numerous research efforts that aimed at detecting melanoma developed machine learning models that outperformed dermatologists.

Although there have been reviews conducted on the use of ML techniques in detecting melanoma and skin cancer [5, 22-24], this study, is the first of its kind to focus specifically on identifying suspicious naevi and UD naevus through the implementation of ML models.

## 6. Conclusion

In summary, this article explores the foundational principles of ML and its potential in assisting with the diagnosis of melanoma as well as the identification of suspicious naevi and UD naevus. Melanoma poses a significant threat as an aggressive type of skin cancer that can rapidly metastasize to lymph nodes. Naevi, benign skin growths consisting of melanocytes or naevus cells, exhibit diverse shapes, sizes, and colours, and the presence of many naevi is associated with an increased risk of melanoma. Early detection of melanoma is crucial due to its propensity for swift and aggressive spread to lymph nodes, sometimes occurring before clinical detection. Therefore, a comprehensive examination of naevi is essential for identifying potential warning signs and implementing timely preventive measures. Despite the wide range of morphological variations observed in naevi, they can be categorized into a limited set of predominant categories or partitions within an individual. Thus, if a naevus does not fit into any of these established categories, it may raise suspicion and be referred to as a UD naevus which could be a sign of melanoma. Although machine learning techniques have been extensively utilized in various healthcare domains, including the study of skin cancer, there is a scarcity of studies specifically focused on applying these techniques to identify suspicious or UD naevi. Further research in this field is necessary to fully harness the potential of machine learning in improving the early detection and management of melanoma and other related skin cancers. This review stands out for its unique focus on suspicious naevi and UD naevus using ML models.